\documentclass[aps,pre,twocolumn,showpacs,superscriptaddress,groupedaddress]{revtex4}  
\pdfoutput=1
\usepackage{graphicx}  
\usepackage{dcolumn}   
\usepackage{bm}        
\usepackage{amssymb}   
\usepackage{color}
\usepackage{ulem}
\newcommand{\eqref}[1]{(\ref{#1})}

\begin{document}



\title{{\color{black}Entropy production in Master Equations and Fokker-Planck Equations: facing the coarse-graining and recovering the information loss}}
\author{D. M. Busiello}
\affiliation{Ecole Polytechnique F\'ed\'erale de Lausanne (EPFL), Institute of Physics Laboratory of Statistical Biophysics, 1015 Lausanne, Switzerland}
\affiliation{Dipartimento di Fisica `G. Galilei', INFN, Universit\'a di Padova, Via Marzolo 8, 35131 Padova, Italy}
\author{A. Maritan}
\affiliation{Dipartimento di Fisica `G. Galilei', INFN, Universit\'a di Padova, Via Marzolo 8, 35131 Padova, Italy}
\date{\today}

\begin{abstract}
Systems operating out of equilibrium exchange energy and matter with the environment, thus producing entropy in their surroundings. Since the entropy production depends on the current flowing throughout the system, its quantification is affected by the level of coarse-graining we adopt. In particular, it has been shown that the description of a system via a Fokker-Planck equation (FPE) lead to an underestimation of the entropy production with respect to the corresponding one in terms of microscopic transition rates. Moreover, such a correction can be derived exactly. Here we review this derivation, generalizing it when different prescriptions to derive the FPE from a Langevin equation are adopted. Then, some open problems about Gaussian transition rates and underdamped limit are discussed. In the second part of the manuscript we present a new approach {\color{black}to dealing with} the discrepancy in entropy production due to the coarse graining by introducing enough microscopic variables to correctly estimate the entropy production within the FPE description. We show that any discrete state system can be described by making explicit the contribution of each microscopic current.
\end{abstract}

\pacs{}
\maketitle

\subsection{E\MakeLowercase{ntropy} P\MakeLowercase{roduction for} \MakeLowercase{coarse-grained} \MakeLowercase{dynamics}}

Different levels of description can be adopted to characterize a physical system, depending on the details we are unaware of or deliberately decided to neglect. In particular, when random variables are introduced to encode the uncontrollable degrees of freedom, the energy balance is properly captured by the concepts proper of the stochastic thermodynamics, when the system is small enough \cite{sekimoto, penr, seif}. In this field, a physical system can be described using two different paradigms: Master Equation (ME) and Fokker-Planck equation (FPE) \cite{gardiner, seif3, threefaces, threefaces2}. The former deals with the probability of occupation of each state $i$ at time $t$, $P_i(t)$, and a set of microscopic transition rates between pair of states $i$ and $j$, {\color{black}$W_{i\rightarrow j}(t)$}. It has the following form:
\begin{equation}
\label{ME}
\dot{P}_i(t) = \sum_j \left( P_j(t)W_{j \rightarrow i}(t)  - P_i(t)W_{i \rightarrow j}(t)  \right)
\end{equation}
which states the the probability to be in the state $i$ evolves according to the balance between the ingoing and the outgoing probability flux of the state itself. Here we assume that $W_{i \rightarrow j}>0$ implies $W_{j \rightarrow i} > 0$, which ensures the ergodicity of the system. {\color{black} Note that here and throughout the manuscript we set $W_{i \to i} = 0$ even though it never enters in the ME above.}

On the other hand, a Fokker-Planck equation described the system via continuous variables by means of a diffusive dynamics:
\begin{equation}
\label{FPE}
\dot{P}(x,t) = - \partial_x \left( A(x) P(x,t) \right) + \partial_x^2 \left( D(x) P(x,t) \right)
\end{equation}
where $P(x,t)$ is the probability to be in the state $x$ at the time $t$, $A(x,t)$ is the drift coefficient and $D(x,t)$ is called the diffusion coefficient. The continuum limit, obtained through a coarse-graining procedure, to derive the latter equation from Eq. \eqref{ME} is called Kramers-Moyal expansion \cite{gardiner}. It relies on the assumption that the first two ``pseudo"-moments \footnote{We called them ``pseudo-moments", since $W(y,r)$ is not a distribution.} of the transition rates are the only ones that are non-negligible, and they are related to $A$ and $D$ respectively.

Although the two descriptions appears be equivalent for certain aspects (e.g. equilibrium properties, dynamics, non-equilibrium steady states), they are not completely {\color{black}interchangeable}, since under coarse-graining some information is lost {\color{black}\cite{espos, hasel, peles, pigo, santi, gomez, nicol, borev}} and some observables might differ in the two descriptions. Among all the possible quantities of this kind, we are particularly interested in the entropy production. The thermodynamic uncertainty relations \cite{dechant, barato} and the fluctuations theorems \cite{lebo, gall, maes, kurch, jarz, crooks, seif2, evans, coll, cil} are examples of fundamental theoretical concepts intimately related to it, which have recently stimulated a lot of experimental invetigations, e.g. estimating the non-equilibrium activity of biological systems \cite{horow} and building efficient molecular engines \cite{how, brown, raz, busie2}.

{\color{black}An enormous amount of studies have investigated how the entropy production gets affected by the coarse-graining \cite{espos, gomez, nicol, borev}. Some of them have also analyzed the possibility to build a consistent diffusive limit starting from a microscopic picture, containing more information about the system \cite{horchem, lebo1, lebo2, muratore}.} In \cite{busie3} it has been shown{\color{black}, in a general setting,} that the entropy production estimated using a diffusive description (FPE) is always less (or equal) to the one estimated starting from a more detailed description (ME). Moreover, this discrepancy can be evaluated explicitly in term of the transition rates. Let us now recall the essential point of this work.

Consider first a system described by Eq. \eqref{ME}. The total entropy production can be estimated using the Schnakenberg's formula \cite{schn, tome, busie}
\begin{equation}
\label{SCHN}
\dot{S}_\mathrm{ME}(t) = \sum_{ij} W_{ij} P_j(t) \log\left(\frac{W_{ij} P_j(t)}{W_{ji} P_i(t)}\right)
\end{equation}
where $W_{ij} \equiv W_{i \rightarrow j}$. Let us assume that the states form an $1D$ lattice of points separated by a distance $\Delta x$. In order to consistently apply the Kramers-Moyal expansion, we {\color{black}introduce} the following form for the transition rates:
\begin{equation}
W_{ij} =\left\{
\begin{array}{ll}
		     W^{(i)}_{\pm k} \delta_{j,i\pm k},&\quad k = 1,...,n \\
                     0&\quad\mathrm{otherwise.}
\end{array}\right.
\end{equation}
with:
\begin{equation}
W^{(i)}_{\pm k} =\Big(1 + \frac{\beta^{(i)}_k \pm \alpha^{(i)}_k}{2}\Delta x \Big)\frac{w^{(i)}_k}{\Delta x^2},
\label{Wpm}
\end{equation}
where $w^{(i)}_k\geq 0$ and $\beta^{(i)}_k\geq |\alpha^{(i)}_k|$ to ensure that $W_{\pm k} \geq 0$ for all $\Delta x$. Notice that we are assuming that, quite generally, the transition rates are state dependent. 

{\color{black}Even if usually only nearest-neighbor jumps are considered, in general, particularly when dealing with system evolving in a generic state space, also long-range interaction can be present \cite{gardiner, schn, montroll, busie, amos, mckane}.}

{\color{black}The proposed transition rates in Eq.~\eqref{Wpm} exhibit a general functional form in order to have a non-zero drift and diffusion coefficient in the diffusive limit, i.e. when $\Delta x \to 0$, and, at the same time, the higher order moments that tend to zero. {\color{black}From a physical perspective, $\Delta x$ can be interpreted as the spacing associated to the spatial discretization {\color{black}\footnote{{\color{black}This form of the transition rates has been also used and justified in \cite{gardiner,horchem}}}. As $\Delta x$ approaches zero, the jumps have to become shorter and more frequent \cite{gardiner}}.} The corresponding FPE has
\begin{eqnarray}
A(x=i \Delta x) &=& \sum_{k=1}^n k\alpha_k^{(i)} w_k^{(i)} \nonumber \\
D(x=i \Delta x) &=& \sum_{k=1}^n k^2 w_k^{(i)}
\end{eqnarray}
Thus performing the diffusive limit, $\Delta x \rightarrow 0$, to Eq. \eqref{SCHN}, and comparing the result with the entropy production of the FPE \cite{pigo2, pre, dechant2}
\begin{eqnarray}
\dot{S}_{\rm FPE}(t) &=& \int dx \frac{\left( A(x) P(x,t) - \partial_x (D(x) P(x,t)) \right)^2}{D(x) P(x,t)} \nonumber \\
&=& \int dx \frac{J(x,t)^2}{D(x) P(x,t)}
\label{Sseif}
\end{eqnarray}
we get the following inequality:
\begin{equation}
 \dot{S}^{\Delta x \rightarrow 0}\equiv \lim_{\Delta x \rightarrow 0}\dot{S}_\mathrm{ME} = \sum_k \int dx \frac{J_k(x,t)^2}{D_k(x) P(x,t)} \geq \dot{S}_{\rm FPE}
\label{Sineqdiscrete}
\end{equation}
Here $J_k(x,t)$ is the flux associated to each microscopic process and $D_k(x,t)$ the corresponding diffusion coefficient, such that $\sum_k J_k(x,t) = J(x,t)$ and $\sum_k D_k(x) = D(x)$, which can expressed in terms of parameters $\alpha$'s and $\beta$'s appearing in eq.\eqref{Wpm}.

{\color{black}The derivation of the inequality in Eq.~\eqref{Sineqdiscrete} follows directly from the Cauchy-Schwarz inequality: $\sum_k a_k^2 \sum_i b_i^2 \geq (\sum_k a_k b_k)^2$ with $a_i=J_i/\sqrt{D_i}$ and $b_i=\sqrt{D_i}$. The equality holds if and only if $a_i$ is proportional to $b_i$ that is $J_i = c D_i$ with  $c$ being a constant. For example this happens when only nearest-neighbor jumps are involved, as in \cite{gingrich}. This conclusion qualitatively agrees with the result found in \cite{muratore} about the protocol minimizing the heat released in a continuous-time Markov process, when the continuum limit is considered.}

Consider now a system with a continuous state space, whose dynamics can be described by a continuous ME:
\begin{equation}
\dot{P}(x,t) = \int dr \left( W(x-r,r) P(x-r,t) - W(x,-r) P(x,t) \right)
\label{ContME}
\end{equation} 
where $W(x,r)$ is the density of the transition rate from the state $x$ to the state $x+r$. {\color{black}In the same spirit as the discrete-case, we introduce the following general form for the transition rates}:
\begin{equation}
W(x,r) = \frac{1}{\epsilon} \frac{1}{\sqrt{d(x) \epsilon}} e^{- f(z(x,r))},
\label{W}
\end{equation}
where $f$ is a generic symmetric function \footnote{The symmetry condition on $f$ ensures that $\langle z \rangle = 0$. This condition is necessary to have the drift coefficient {\color{black}of the corresponding FPE} of order $1$.} and $z(x,r) = (r - A(x) \epsilon)/\sqrt{d(x) \epsilon}$. 

{\color{black} Eq.~\eqref{W} can be seen as the continuous-state counterpart of a discrete-state system with a continuous jump distribution. The term $A(x) \epsilon$ represents the deterministic part of the jump whereas $\sqrt{d(x) \epsilon}$ is the amplitude of the random noise.}

{\color{black}The functional form in Eq. \eqref{W} is also suitable for the Kramers-Moyal expansion, i.e. the first two ``pseudo"-moments are the only ones different from zero in the diffusive limit, i.e. $\epsilon \to 0$. Indeed, the scaling introduced by the parameter $\epsilon$ has been chosen so that this latter condition holds. {\color{black}The physical interpretation of $\epsilon$ can be understood when $f(z) = z^2$, i.e. for a Gaussian form of the transition rates. In this case, since $W(x,r)$ should correspond to the short-time propagator divided by the time spacing $\Delta t$ \cite{gardiner}, we can obtain immediately this result by identifying $\epsilon = \Delta t$. This means that, in this case, $\epsilon$ has to be related to the temporal coarse-graining we are performing on the dynamics} {\color{black}\footnote{{\color{black}The scaling of the model parameters is an unvoidable consequence of the coarse-graining procedure, in the same spirit of what has been shown in the context of the renormalization group \cite{wilson}.}}}.} Without loss of generality we have chosen $f(0)$ such that $\int dz  e^{-f(z)} = 1$. {\color{black}Then,} the corresponding drift and diffusion coefficient {\color{black}are} $A(x)$ and $D(x) = (1/2) d(x) \langle z^2 \rangle$, respectively.

Performing the diffusive limit, we end up with the following entropy production:
\begin{widetext}
\begin{eqnarray}
\dot{S}^{\epsilon\rightarrow 0}\equiv \lim_{\epsilon\rightarrow 0}\dot{S}_{\mathrm{ME}}(t) &=& \int dy \frac{\left( A(y) P(y) - \partial_y \left( D(y) P(y) \right)\right)^2}{D(y) P(y)} + \nonumber \\
&+& \left( \langle z^2 \rangle \langle (\partial_z f(z))^2 \rangle - 1 \right) \int dy \frac{A(y)^2}{D(y)} P(y) + \nonumber \\
&+& \left( 3 - \langle z^2 (\partial_z f(z))^2 \rangle \right) \int dy \frac{A(y) \partial_y D(y)}{D(y)} P(y) + \nonumber \\
&+& \frac{1}{4} \left( - 9 + \frac{\langle z^4 (\partial_z f(z))^2 \rangle}{\langle z^2 \rangle} \right) \int dy \frac{(\partial_y D(y))^2}{D(y)} P(y)
\label{Entgen}
\end{eqnarray}
\end{widetext}
where $\langle\cdot\rangle = \int dz \cdot e^{-f(z)}$. 
{\color{black} Because of the diffusive limit the only surviving terms - the ones reported in Eq. \eqref{Entgen} - are of order $\epsilon^0$, as the drift and the diffusion coefficient in the corresponding FPE. 

The detailed derivation of this formula is presented in \citep{busie3}: the entropy production for the ME is evaluated via the Schnakenberg's formula \cite{schn}, and then the diffusive limit is performed. This allows to get the entropy production, Eq. \eqref{Entgen}, without relying directly on the expression of the FPE, which would lead to a different expression for it, $\dot{S}_{\rm FPE}$ (see beolw). We can see, in fact, that the amount of information loss does not vanish in the diffusive limit, as it is evident from the last three lines in Eq.~\eqref{Entgen}. This picture emerges also at the level of discrete-state systems, as it was seen in Eq.~\eqref{Sineqdiscrete}, for which a naive interpretation in terms of microscopic currents is clear and intuitive.}

Since the general formula is quite cumbersome, it is useful to restrict ourselves to some simple yet interesting cases:
\begin{itemize}
\item Constant diffusion $D$:
\begin{equation}
\dot{S}^{\epsilon\rightarrow 0} = \dot{S}_{\rm FPE} + \left( \langle z^2 \rangle \langle (\partial_z f)^2 \rangle - 1 \right) \int dy \frac{A^2}{D} P
\end{equation}
Notice that the correction is always positive (it is enough to apply the Cauchy-Schwartz inequality to $1= \langle z\partial_z f(z) \rangle^2$) except when $W(x,r)$ is Gaussian, i.e. $f(z) -f(0) \propto z^2$.
\item Spatial dependent diffusion $D(x)$ and zero drift $A(x) = 0$:
\begin{equation}
\dot{S}^{\epsilon\rightarrow 0} = \dot{S}_{\rm FPE} + \frac{1}{4} \left( \frac{\langle z^4 (\partial_z f)^2 \rangle}{\langle z^2 \rangle}-9 \right) \int dy \frac{(\partial_y D)^2}{D} P
\end{equation}
Also in this case it is possible to show that the correction is positive \cite{busie3}.
\item Non-constant diffusion $D(x)$ and Gaussian transition rates:
\begin{equation}
\dot{S}^{\epsilon\rightarrow 0} = \dot{S}_{\rm FPE} + \frac{3}{2} \int dy \frac{(\partial_y D)^2}{D} P
\label{GaussDx}
\end{equation}
\end{itemize}

These are the main results of \cite{busie3}. At this point we can state that, in order to have a consistency between ME {\color{black}and} FPE in terms of entropy production, the transition rate density has to be Gaussian, and the diffusion has to be space independent.

\subsubsection{From Ito to Stratonovich prescription}

The FPE can be also derived from a suitable Langevin equation. Since the transition rates $W(x,r)$ in Eq. \eqref{W} have been evaluated at the starting position $x$, the Langevin equation corresponds to the Ito prescription. As mentioned above, the FPE associated to the transition rates in Eq. \eqref{W} is:
\begin{equation}
\dot{P}(x,t) = -\partial_x \left( A(x) P(x,t) \right) + \frac{1}{2} \partial_x^2 \left( 2 D(x) P(x,t) \right)
\label{FPEITO}
\end{equation} 
whereas the corresponding Langevin equation turns out to be
\begin{equation}
\dot{x}(t) = A(x) + \sqrt{2 D(x)} \xi(t),
\label{Ito}
\end{equation}
with the Ito prescription. The whole machinery presented in the previous section can be nevertheless extended when the Stratonovich prescription is adopted. To this aim, we introduce the mid-point rule in the definition of the transition rates as follows:
\begin{equation}
W(x,r) = \frac{1}{\epsilon} \frac{1}{\sqrt{d(x+r/2) \epsilon}} e^{- f(z(x+r/2,r))}
\label{Wmid}
\end{equation}
Expanding Eq. \eqref{Wmid} in powers of $r$ one can derive the corresponding FPE as before \footnote{the method to carry on the analytic calculations are the ones adopted in \cite{busie3}, so we refer to this for a detailed discussions.}, obtaining:
\begin{equation}
\dot{P}(x,t) = -\partial_x \left( A_{\rm Strat}(x) P(x,t) \right) + \frac{1}{2} \partial_x^2 \left( 2 D(x) P(x,t) \right)
\label{FPEITO}
\end{equation}
where:
\begin{equation}
A_{\rm Strat}(x) = A(x) - \frac{1}{2} \partial_x D(x)
\end{equation}
As expected, this FPE corresponds to the same Langevin equation, Eq.~\eqref{Ito}, interpreted according to the Stratonovich prescription.

Performing the Kramers-Moyal expansion on Eq. \eqref{Wmid}, we can estimate the entropy production of the system of interest when it is described using the Stratonovich prescription, $\dot{S}_{\rm Strat}^{\epsilon \rightarrow 0}$. Up to the zeroth order in $\epsilon$, we end up with the following expression:
\begin{equation}
\dot{S}_{\rm Strat}^{\epsilon \rightarrow 0} = \dot{S}^{\epsilon \rightarrow 0}\big|_{A(x)\Rightarrow A_{\rm Strat}(x)}
\end{equation}
It is worth noting that our procedure to implement a given prescription from a microscopic perspetive is consistent, since it resembles the entropy production obtained just changing prescription at the level of the Langevin equation. The coarse-graining corrections are the same discussed before. Clearly the difference between the Ito and the Stratonovich prescription appears only when we have a space-dependent diffusion.

In general, different prescriptions are possible to interpret a Langevin equation \cite{lau, gardiner}, each one affecting the value of the entropy production. The general form of the transition rates is:
\begin{equation}
W(y,r) = \frac{1}{\epsilon} \frac{1}{\sqrt{d(x+\lambda r) \epsilon}} e^{-f(z(x+\lambda r,r))}
\end{equation}
This means that we have to evaluate $d(x)$ and $A(x)$ at an intermediate point between $x$ and $x+r$ when performing the diffusive limit. The choice $\lambda=0$ corresponds to the Ito prescription, while $\lambda = 1/2$ to the Stratonovich one. The resulting entropy production in this more general case is simply:
\begin{equation}
\dot{S}_{\lambda}^{\epsilon \rightarrow 0} = \dot{S}^{\epsilon \rightarrow 0}\big|_{A(x)\Rightarrow A(x) - \lambda \partial_x D(x)}
\label{Slambda}
\end{equation}
Using Eq. \eqref{Slambda} and Eq. \eqref{GaussDx}, and considering the case of Gaussian transition rates, it is clear that does not exist a prescription, i.e. a choice of $\lambda \in [-1,1]$, such that the correction terms due to the coarse-graining vanish.

\subsubsection{Open problems: Gaussian transition rates and underdamping}

The problem of the coarse-graining procedure leaves behind some open questions which we think deserve a discussion.

The first one concerns the particular case of Gaussian transition rates. In fact, we have seen that, when the diffusion does not depend on space, having Gaussian transition rates $W(y,r)$ leads to a null correction to the entropy production with respect to the one obtained directly from the FPE. In other words, this is the only case in which we have no loss of information during the coarse-graining procedure. We guess that this result is related to the fact that, in order to describe a microscopic dynamics as a Fokker-Planck equation, one needs to assume a Gaussian white noise, in order to prevent the onset of inconsistencies in non-equilibrium quantities \cite{mazur, mazur2}.

As a second point, we want to discuss the applicability of our result. First of all notice that our formalism can be consistently applied also when the FPE is used to describe an evolution over an abstract state space. In particular, interacting ecological and social systems (e.g. bacteria, species, humans)  are examples for which both the microscopic descriptions, in terms of the ME and the FPE formalism, are typically used \cite{amos, mckane}. Thus, a thermodynamic (or information-theoretic) approach to population dynamics could lead to a better evaluation of the non-equilibrium activity of the system and a more careful identification of the physical quantities in play (e.g. energy, heat dissipated).

However, one of the most straightforward application is related to the diffusion of a particle in the real space. In this case we know that the description in term of an overdamped process (1D FPE) is not able to capture all the essential physical features of the system of interest {\color{black}\cite{celani, pre, muratore2}} when a space-dependent diffusion in present, $D(x)$. It will be therefore interesting to generalize our approach to the underdamped case, a 2D FPE in the position-velocity space, and we will investigate this topic in future works.

\subsection{C\MakeLowercase{oarse-graining without information loss}}

Since the ME describes a process from a microscopic perspective, it is not obvious that a specific form of the transition rates exists such that the corresponding FPE and respective entropy production exhibit no coarse-graining corrections. In fact, it is easy to see that{\color{black}, when the diffusion coefficient depends on space, the corrective term in the last line of Eq.~\eqref{Entgen} does not {\color{black} vanish} even for Gaussian transition rates (see Eq.~\eqref{GaussDx}), for which all the other corrections vanish. In fact, it is possible to show that {\color{black}it is always positive \cite{busie3}}} for any choice of $W(y,r)$ that is consistent with the Kramers-Moyal expansion.

On the contrary, it is a well-posed question to ask whether it is possible to introduce a new coarse-graining procedure, other than the standard Kramers-Moyal, so that the least information is lost in the diffusive limit. In what follows, we present {\color{black}a possible approach} in this direction.

\subsubsection{Dynamics of the microscopic processes}

We deal with this problem in the case of a discrete state system whose dynamics is described by Eq. \eqref{Wpm}. {\color{black}As we have shown in Eq.~\eqref{Sineqdiscrete}, the excess entropy production with respect to the one obtained directly from the FPE is due to the presence of different channels through which the system can jump. In other words, instead of using a coarse-grained current $J = \sum_k J_k$, the Schnakenberg's entropy production keeps track of all the microscopic currents separately. Here we decompose the ME into a dynamical evolution of different state subspaces, where the information about the microscopic currents is retained also in the diffusive limit. This procedure allows us to reconstruct the correct entropy production starting from a new set of coupled diffusive equations.}

Consider first the simplest case $k=1,2$, i.e. a random walk with jumps only to the first and second-nearest neighbors. For sake of simplicity, but without loss of generality, we restrict ourselves to the case in which the transition rates does not depend on the state $i$. Let us write the ME for the probability to be in the odd, $P_{2i-1}$, and even, $P_{2i}$, states:
\begin{eqnarray}
\dot{P}_{2i} &=& W_{+1} P_{2i-1} + W_{-1}P_{2i+1} - (W_{+1}+W_{-1})P_{2i} + \nonumber \\
&+& W_{+2}P_{2i-2} + W_{-2}P_{2i+2} - (W_{+2}+W_{-2})P_{2i} \nonumber \\
\dot{P}_{2i-1} &=& W_{+1} P_{2i-2} + W_{-1}P_{2i} - (W_{+1}+W_{-1})P_{2i-1} + \nonumber \\
&+& W_{+2}P_{2i-3} + W_{-2}P_{2i+1} - (W_{+2}+W_{-2})P_{2i-1} \nonumber \\
\label{Pevenodd}
\end{eqnarray}
If the Kramers-Moyal expansion were performed on both equations, it would lead to two identical FPE, since, in the diffusive limit, we have no distinction between odd and even states. For sake of simplicity, let us write such a diffusion equation through its operator: 
\begin{equation}
\dot{P} = \mathcal{L}(A,D) P = \left( - A \partial_x + \frac{1}{2} D \partial_x^2 \right) P(x)
\end{equation}
where:
\begin{eqnarray}
A &=& \lim_{\Delta x \to 0} \Delta x \sum_{k=-n}^n k W_{k} \nonumber \\
D &=& \lim_{\Delta x \to 0} (\Delta x)^2 \sum_{k=-n}^n k^2 W_{k}
\end{eqnarray}
However a Fokker-Planck operator can be associated to each microscopic process: the one involving jump of size $k$ will be denote by $\mathcal{L}_k$, such that:
\begin{equation}
\mathcal{L}(A,D) = \sum_{k=1}^n \mathcal{L}_k
\end{equation}
where:
\begin{eqnarray}
\mathcal{L}_k &=& - A_k \partial_x + \frac{1}{2} D_k \partial_x^2 \nonumber \\
A_k &=& \lim_{\Delta x \to 0} \Delta x ~k \left( W_{+k} - W_{-k} \right) \nonumber \\
D_k &=& \lim_{\Delta x \to 0} (\Delta x)^2 ~k^2 \left( W_{+k} + W_{-k} \right)
\end{eqnarray} 
In this simple situation, $n=2$.

{\color{black}We want to point out that the splitting of the FPE operator $\mathcal{L}$ into the subset $\{\mathcal{L}_k\}_{k=1,...n}$ is a natural choice when jumps of different size are involved. Moreover, the form of $A_k$ and $D_k$ coincides with the one that appears in Eq.~\eqref{Sineqdiscrete}, such that $\sum_k A_k = A$ and $\sum_k D_k = D$.}

Alternatively, we can consider $P_{2i}$ and $P_{2i-1}$ as two independent quantities. We recall that, in the diffusive limit, $i=x/\Delta x$, $P_{2i}= P_1 (x)\Delta x$, and $P_{2i-1}= P_2 (x)\Delta x$. In order to formulate the system of Eq.s \eqref{Pevenodd} as a function of $P_1$ and $P_2$ only, we expand all the terms involved in the process with $k=1$ for $P_1$ around $P_2$, and viceversa for $P_2$, obtaining:
\begin{widetext}
\begin{eqnarray}
\dot{P}_1 &=& W_{+1} P_2 + W_{-1}(P_2 + 2 \Delta x \partial_x P_2 + 2 \Delta x^2 \partial_x^2 P_2) + \nonumber \\
&-& (W_{+1}+W_{-1})(P_2 + \Delta x \partial_x P_2 + \frac{1}{2} \Delta x^2 \partial_x^2 P_2) + \mathcal{L}_2 P_1 = \mathcal{L}_1 P_2 + \mathcal{L}_2 P_1 \nonumber \\
\dot{P}_2 &=& W_{-1} P_1 + W_{+1}(P_1 - 2 \Delta x \partial_x P_1 + 2 \Delta x^2 \partial_x^2 P_1) + \nonumber \\
&-& (W_{+1}+W_{-1})(P_1 - \Delta x \partial_x P_1 + \frac{1}{2} \Delta x^2 \partial_x^2 P_1) + \mathcal{L}_2 P_2 = \mathcal{L}_1 P_1 + \mathcal{L}_2 P_2 \nonumber \\
\label{Pplusminus}
\end{eqnarray}
\end{widetext}
where the last equalitites have been derived by expliciting the form of the transition rates and performing the limit $\Delta x \rightarrow 0$. Notice that {\color{black}we have derived two coupled diffusive equations involving the operators that govern the microscopic processes, $\mathcal{L}_1$ and $\mathcal{L}_2$ in this simple case. As a consequence, the system is described in terms of four parameters, two microscopic drift and two diffusion coefficients. Indeed, these latter are needed to match the correct entropy production directly from the diffusive description in Eq.~\eqref{Pplusminus}, noting that $P_1 + P_2$ evolves according to the standard FPE.}

It is straightforward to generalize this procedure for processes with jumps of size $k=1,...n$. Thus, we end up with $n$ different differential equations, for $P_1,...P_n$, depending on $2n$ parameters, $A_k$ and $D_k$, for $k=1,...n$. However, notice that{\color{black}, also in the general case,} only the sum of all the probabilities, $\Sigma P = \sum_{j=1}^n P_j$, satisfies the standard FPE of the system. It is then immediate to see that{\color{black}, again,} the entropy production in Eq. \eqref{Sineqdiscrete} can be estimated from the set of parameters $\{ A_k, D_k \}_{k=1,...n}$ and $\Sigma P$.

\subsection{Conclusions}

In the present manuscript we have discussed how the
entropy production gets affected by the coarse-graining
that we perform on the dynamics of a stochastic system.
As expected, since in passing from a ME to a FPE we
neglect some information, the entropy production will be
underestimated when directly evaluated within a diffusive formalism.

We have generalized this derivation to the case in
which a different prescription is used to describe the system in terms of a Langevin equation. This completes the
formulation in a consistent way, involving all the frameworks commonly used to characterize a stochastic system, ME, FPE and stochastic differential equations.

In the last {\color{black}part} we propose a novel approach to the coarse-graining of a microscopic system, which, in principle, is
able to capture all the information that are neglected performing the standard Kramers-Moyal expansion. This
result also suggest that the source of discrepancy comes
from the microscopic fluxes flowing through the system,
which are averaged in a diffusive description. The experimental work of Battle et al. [45] can be considered as
an important step in this direction, since it quantifies the
breakage of detailed balance due to microscopic fluxes in
a specific system operating out of equilibrium.

\subsection{Acknowledgements}

We acknowledge U. Seifert, C. Jarzynski, C. Maes, J. Hidalgo and S. Suweis for useful comments and discussions. {\color{black} A.M. was supported by ``Excellence Project 2017" of the Cariparo Foundation.}

\end{document}